\documentclass[amssymb,twocolumn,
aps,floatfix,showpacs]{revtex4}
\usepackage{graphicx}

\begin{document}
\title{Interference between competing pathways in atomic harmonic generation}
\author{A.C. Brown\textsuperscript{*}}
\author{S. Hutchinson\textsuperscript{*}}
\author{M. A. Lysaght\textsuperscript{+}}
\author{H. W. van der Hart\textsuperscript{*}}
\address{\textsuperscript{*}Centre for Theoretical Atomic, Molecular and Optical Physics,
Queen's University Belfast, Belfast, BT7 1NN, UK.\\
\textsuperscript{+}Irish Centre for High End Computing, Tower Building, Grand Central Quay, Dublin 2, Ireland.}

\date{}

\begin{abstract} 
We investigate the influence of the autoionizing $3s3p^6n\ell$ resonances on
the fifth harmonic generated by 200 - 240 nm laser fields interacting with Ar.
To determine the influence of a multi-electron response we have developed
capability within time dependent $R$-matrix theory to determine the harmonic
spectra generated. The fifth harmonic is affected by interference between the
response of a $3s$ electron and the response of a $3p$ electron, as
demonstrated by the asymmetric profiles in the harmonic yields as functions of
wavelength.  
\end{abstract}


\maketitle

%%%%	Introduction	%%%

Harmonic generation (HG) is one of the fundamental processes that can occur in
light-matter interaction.  The process is of strong technological interest,
since it lies at the heart of ultra-short light pulse generation
\cite{paul2001}. However, HG can also be used as a measurement tool, for
example, to study molecular changes on the sub-femtosecond timescale
\cite{baker2006}. Recent articles have postulated various other applications of
high harmonics, including electron interferometry and the tomographic imaging
of electron orbitals \cite{corkum2011}.  HG has also been used experimentally
to demonstrate the importance of electron-correlation in ultra-fast molecular
dynamics \cite{smirnova2009} and multi-electron dynamics in
atomic systems \cite{shiner2011}.

In order to probe the influence of multi-electron dynamics on HG 
theoretically, we find it more expedient to consider atomic systems.
Although detailed calculations for molecules are possible using sophisticated
codes, it is much more difficult to identify the multi-electron dynamics within
a molecular system than within an atomic system. Even so, many methods
developed for the determination of harmonic spectra in atoms rely on the single
active electron approximation \cite{Gavrila,Schafer1997}. Several methods have
been developed for the description of two-electron systems \cite{Parker,
Lambropoulos}, but these methods are not easy to extend to general atoms. New
methods are thus needed to gain insight into the influence of multi-electron
dynamics on harmonic spectra in general multi-electron atoms. 

At Queen's University Belfast we have developed time-dependent {\it R}-matrix
(TDRM) theory over the last five years expressly to study the behavior of
multi-electron atoms in short, intense light fields \cite{tdrm}. The advantage
of this method in studying multi-electron dynamics was shown in a pump-probe
study of C$^+$ \cite{Lys09b}: by varying the delay between the pump and probe
pulses, oscillations in the total ionization yield were observed, which could
be linked to the correlated motion of two electrons. Because of the interest in
using harmonic spectra to observe multi-electron dynamics, we have developed
capability within the time-dependent {\it R}-matrix approach to determine
harmonic spectra for a single atom.

\begin{figure}
\begin{centering}
\includegraphics[width=7.9cm]{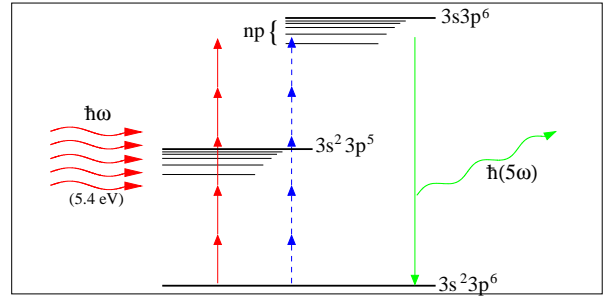}
\caption{(Color online) The harmonic yield in Argon is determined by two
competing pathways:  absorption of at least three photons can excite a $3p$
electron (red solid line) into the continuum, while absorption of at least five
photons can excite a $3s$ (blue dashed line) electron to the $np$ Rydberg
series converging onto the $3s3p^6$ threshold.  \label{fig:energy}}
\end{centering}
\end{figure}

As a first demonstration of this capability, we report on the influence of the
$3s3p^6n\ell$ Rydberg series on single atom harmonic spectra in Ar. In
particular, we focus on the wavelength range for which the fifth harmonic
coincides with these resonances: 200 - 240 nm. Recent studies have focused on
the parallels between photoionization and HG \cite{worner2009,higuet2011},
where, in particular, the Cooper minimum in Ar was investigated.  However,
minima in the photoionization cross section can also occur near window
resonances, where resonances actually lead to a decrease in the photoionization
cross section. The $3s3p^6n\ell$ resonances are examples of window resonances
\cite{madden69}. Their decay towards $3s^23p^5\varepsilon\ell'$ continua
requires a good description of dielectronic repulsion. The harmonic yield will
be determined by two interfering pathways, (Fig. \ref{fig:energy}): excitation of a
$3s$ electron to $np$, and excitation of a $3p$ electron into the continuum. A
multi-electron calculation is essential to describe the harmonic yield
accurately. This study thus provides a good test case.

%%%%	TDRM Theory	%%,%
To describe the extension to compute a system's harmonic response, we
first give a brief overview of TDRM theory.  A more thorough discussion can be
found in \cite{tdrm}.  The time-dependent Schr\"{o}dinger equation for an atom
containing ($N+1$) electrons, and nuclear charge $Z$, is 
\begin{equation}
\label{tdse}
i\frac{\partial}{\partial t} \Psi \left( \mathbf{X}_{N+1},t\right)= 
H\left(t\right)\Psi\left(\mathbf{X}_{N+1},t\right).
\end{equation}
The Hamiltonian, $H$, contains both the non-relativistic Hamiltonian of the
$N+1$-electron atom or ion in the absence of the laser field and the laser
interaction term.  The laser field is described using the dipole approximation
in the length form, and is assumed to be linearly polarized and spatially
homogeneous. This form provides the most reliable ionization
yields when only a limited amount of atomic structure is included
\cite{hutch2010}.

To propagate the wave function in time, time is discretized into
steps of $\Delta t$ such that
\begin{equation}
\label{timestep}
t_q=q\Delta t, \mbox{\hspace{0.5cm}} q=0,1,2,\ldots.
\end{equation}
Then, using the unitary Cayley form of the time evolution operator,
$\exp\left[-itH\left(t\right)\right]$, the wavefunction $\Psi$ in Eq. (\ref{tdse})
at $t_{q+1}$ can be written in terms of the wavefunction at $t_q$
\begin{equation}
\label{recurs1}
\left[H\left(t_{q+1/2}\right)-E\right]\Psi\left(\mathbf{X}_{N+1},t_{q+1}\right)=
\mathbf{\Theta}\left(\mathbf{X}_{N+1},t_q\right),
\end{equation}
where
\begin{equation}
\label{recurs2}
\mathbf{\Theta}\left(\mathbf{X}_{N+1},t_q\right)=
-\left[H\left(t_{q+1/2}\right)+E\right]\Psi\left(\mathbf{X}_{N+1},t_q\right),
\end{equation}
and $H\left(t_{q+1/2}\right)$ is the time-dependent Hamiltonian at the
midpoint. $E$ is given by $2i/\Delta t$.  In equations (\ref{recurs1}) and
(\ref{recurs2})
$\mathbf{X}_{N+1}=\mathbf{r}_1\sigma_1,\mathbf{r}_2\sigma_2,\cdots,\mathbf{r}_{N+1}\sigma_{N+1}$
where $\mathbf{r}_i\sigma_i$ are the space and spin coordinates of the
$i$\textsuperscript{th} electron.  If the wavefunction at time $t=t_q$ is
known, then Eq. (\ref{recurs1}) allows us to propagate the solution forward in
time.

In $R$-matrix theory, configuration space is partitioned into two regions: an
inner and an outer region. In the inner region, all electrons are within a
distance $a_{\mathrm{in}}$ of the nucleus, and all interactions between all
electrons are taken into account. In the outer region, one electron has moved
beyond a distance $a_{\mathrm{in}}$, and exchange interactions between this
electron and the electrons remaining close to the nucleus can be neglected. The
outer electron thus experiences only the long-range multipole potential of the
residual $N$-electron core and the laser field.

In the outer region, $a_{\mathrm{in}}\le r_{N+1}\le a_{\mathrm{out}}$, we have
\begin{eqnarray}
\label{outexpand}
\lefteqn{ \Psi\left(\mathbf{X}_{N+1},t_{q+1}\right)= } \nonumber \\
& \sum\limits^n_{p=1}\bar{\Phi}^{\gamma}_p\left(\mathbf{X}_N;
\hat{\mathbf{r}}_{N+1}\sigma_{N+1}\right)r^{-1}_{N+1}F_p\left(r_{N+1}\right),
\end{eqnarray}
where $\bar{\Phi}^{\gamma}_p$ are channel functions which couple possible final
states of the Ar\textsuperscript{+} ion with spin and angular momentum of the
ejected electron.  The $F_p$ are the radial wavefunctions which describe the
motion of the ejected electron in the $p$\textsuperscript{th} channel, where
$p$ is a channel index.

We can now follow $R$-matrix procedures \cite{vdh07} to evaluate
Eq. (\ref{recurs1}) at the boundary $a_{\mathrm{in}}$ as a
matrix equation
\begin{equation}
\label{frtmat}
\mathbf{F}\left(a_{\mathrm{in}}\right)=
\mathbf{R}(a_{\mathrm{in}})\bar{\mathbf{F}}\left(a_{\mathrm{in}}\right)+\mathbf{T}\left(a_{\mathrm{in}}\right),
\end{equation}
in which the wavefunction at the boundary is described in terms of its
derivative, $\bar{\mathbf F}$, plus an inhomogeneous vector, ${\mathbf T}$,
arising from the $\Theta$ term in Eq. (\ref{recurs1}).  Equation (\ref{frtmat})
connects the inner and outer region wavefunction at the boundary
$a_{\mathrm{in}}$. 

Given an initial inner region wavefunction, $\mathbf{R}$ and $\mathbf{T}$ are
evaluated  at the boundary. Subsequently, they are propagated outwards in space
up to a boundary, $a_{\mathrm{out}}$ where it can be assumed that the
wavefunction $\mathbf{F}$ has vanished. The wavefunction vector $F$ is set to
zero and then propagated inwards to the inner region boundary. Once the
$F$-vector has been determined at each boundary point, the full wavefunction
can be extracted from the $R$-matrix equations. Equation (\ref{recurs1}) then
allows us to step this solution forward in time.

%%%%	HG Theory	%%%
HG arises from the oscillating dipole moment induced by the laser field. This
is given by the expectation value of the electric dipole operator,
\begin{equation}
\mathbf{d}\left(t\right)=\langle \Psi \left(t\right)
|-e\mathbf{Z}|\Psi\left(t\right)\rangle,
\label{inducedip}
\end{equation}
where $e$ is the electronic charge and {\bf Z} the total position operator
along the laser polarization axis.  The power spectrum of the emitted radiation
is then given by the Fourier transform of either the dipole moment, its
velocity or acceleration \cite{madsen2011}.  Many techniques involving high
order harmonics use the acceleration representation as it gives better
resolution for the high-order peaks \cite{dipacc}. However, this has
predominantly been verified for single-active-electron systems, whereas in the
multi-electron TDRM method the length form of the dipole operator has proven to
be far more reliable \cite{hutch2010}. 

%%%%	Results	%%%
To describe Ar, we use the $R$-matrix basis developed for single
photon-ionization \cite{taylor1975} with an inner region radius of 20 a.u. The
set of continuum orbitals contains 60 $B$-splines for each angular momentum,
$\ell$, of the continuum electron. We use both the $3s^23p^5 \mbox{ }^2P^o$ and
$3s3p^6 \mbox{ }^1P^o$ states of Ar\textsuperscript{+} as target states.  The
description of Ar includes all $3s^23p^5\epsilon\ell$ and $3s3p^6\epsilon\ell$
channels up to $L_{max}=15$. In the outer region we propagate $\mathbf{R}$ and
$\mathbf{T}$ out to a radial distance of 1600 a.u. to prevent any
reflections of the wavefunction from the outer region boundary.  The outer
region is divided into sectors of width 2 a.u. and each sector contains 35
$B$-splines of order $k=9$ per channel.  We use similar pulse shapes to those
used in \cite{tdrm}: a 3 cycle $\sin^2$ turn on/off for the electric field.
Peak intensity is set at $2\times10^{13}\mbox{ }\mathrm{Wcm}^{-2}$ with varying
numbers of cycles at peak intensity; the time step is 0.1 a.u.
 
\begin{figure}
\begin{center}
\includegraphics[width=7.9cm]{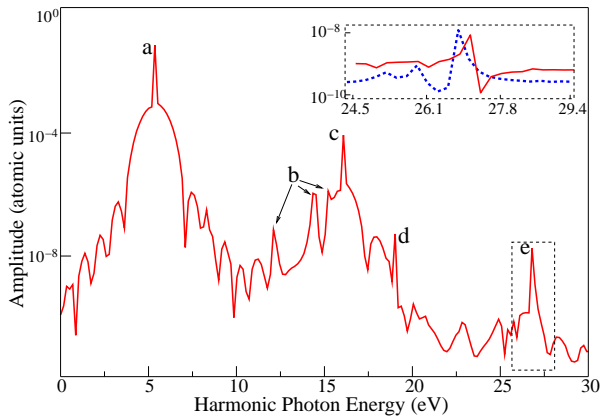}
\end{center}
\vspace{-8pt}  
\caption{(Color online) The power spectrum produced by a 25 cycle,
$2\times10^{13}\mbox{ }\mathrm{Wcm}^{-2}$ laser pulse of photon energy 
5.4eV. The features labelled are (a) First harmonic peak,
(b) Effects due to $3s^23p^5n\ell$ Rydberg series,
(c) Third harmonic peak,
(d) Peak due to $3s^23p^54p$ states plus a photon, and
(e) Fifth harmonic peak.
Inset:  The fifth harmonic region of the power spectra for
two 25 cycle pulses of photon energy 5.36 eV (blue, dashed line), and 5.42 eV 
(red, solid line). The
change in the profile of the peaks is typical of the presence of a resonance
in between these two frequencies.\label{fig:spectra}}
\end{figure}

Figure \ref{fig:spectra} shows a typical power spectrum obtained for an
incident pulse with a photon energy of 5.4 eV and 25 cycles at peak intensity.
As well as the three harmonic peaks, several other features are visible. The
sideband structure of the harmonic peaks is due to the turn on/off of the laser
pulse. In the photon energy region between 12.1-15.2 eV, several peaks are
observed arising from the $3s^23p^5n\ell$ Rydberg series: $4s$ at 12.1 eV, the
$5s/3d$ manifold at 14.5 eV and the $4d/6s$ at 15.1 eV. Higher manifolds are
not separated at this pulse length.  The third harmonic peak is clearly
visible; it lies just above the $3s^23p^5$ threshold of Ar$^+$. Its proximity
to the threshold may mean that it still overlaps the Rydberg series leading up
to this threshold. The peak at 18.8 eV cannot be associated directly with any
atomic structure but this energy corresponds to the energy of $3s^23p^54p$ plus
one 5.4 eV photon. Finally, the fifth harmonic is visible at an energy of 26.9
eV.

The power spectrum was compared with calculations carried out
using the {\it R}-matrix Floquet (RMF) method \cite{plummer2002}.  Good
agreement was found at 248nm for the first harmonic rate at intensities
between 6.5 and $8.5\times10^{12}\mbox{ }\mathrm{Wcm}^{-2}$.  The RMF
calculations also obtained third and fifth harmonic rates within the $3d/5s$
manifold. In RMF theory these states are separate, but this requires unfeasibly
long pulses in TDRM theory. Thus no comparison is possible beyond the first
harmonic.

\begin{figure}
\centering
\includegraphics[width=7.8cm]{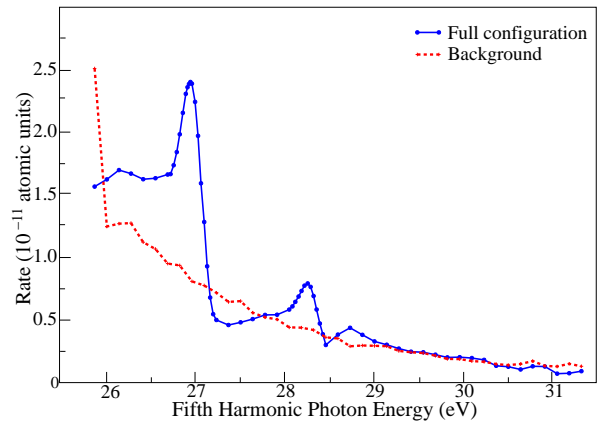}
\caption{(Color online) The fifth harmonic rate for Ar in a 25 cycle pulse of
varying frequency, for a target with the full configuration (blue, solid line)
and one without any contribution from the $3s3p^6n/\varepsilon\ell$ series
(red, dashed line). The resonances in the fifth harmonic rate correspond to the
$3s3p^64p$ (26.9eV) and $3s3p^65p$ (28.3 eV) resonances. \label{fig:res_spec}}
\end{figure}

The main interest in the present manuscript is the fifth harmonic, which is
strongly influenced by the $3s3p^6n\ell$ Rydberg series. Photoionization
experiments and calculations have demonstrated that the Rydberg series appears
as window resonances \cite{madden69,taylor1975}. 
Figure \ref{fig:spectra} also shows part of the power spectra at two distinct
photo energies, 5.36 eV and 5.42 eV.  The resonance profile in the harmonic
spectrum changes significantly. This is typical when a resonance lies in between
these two photon energies. In order to assess these changes,
the fifth harmonic rate is determined over a range of photon energies.
For each photon energy, we integrate the harmonic rate across the bandwidth of
the harmonic peak. This approach yields the harmonic rates shown in Fig.
\ref{fig:res_spec}. We can see resonances occurring at energies corresponding to
the $3s3p^64p$ and $3s3p^65p$ states.  Additional resonances may be present at
higher photon energies, but the energy uncertainty makes their assessment
difficult.

\begin{figure}[t]
\centering
\includegraphics[width=7.5cm]{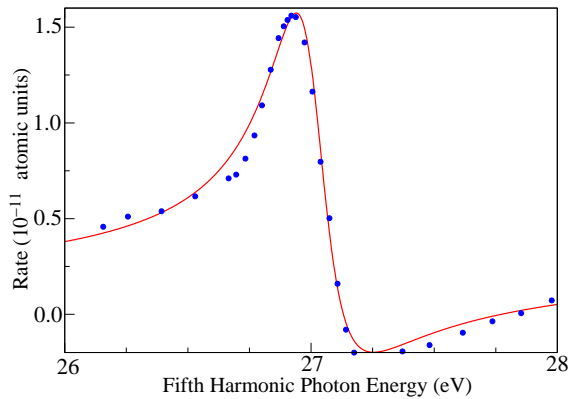}
\caption{(Color online) The fifth harmonic resonance profile for Ar in a
25 cycle laser pulse of intensity $2\times10^{13}\mbox{ }\mathrm{Wcm}^{-2}$.
The data (blue points) is fitted by a Fano profile (red line) with asymmetry
parameter, $q=-1.84$ and line width $\Gamma=0.27$ eV,
and a linear background term. \label{fig:res_prof}}
\end{figure}

In order to obtain resonance parameters, calculations are also performed with a
reduced basis: a $3s^23p^5$ single-target-state description of Ar$^+$, and an
Ar basis which contains no $3s3p^6n/\varepsilon\ell$ functions. This basis
produces a smooth spectrum with a rate which can act as a background.
Subtraction of this background rate then reveals a resonance profile due to the
$3s3p^6n\ell$ Rydberg series, from which resonance parameters can be
established. The resonance profile in Fig. \ref{fig:res_prof} was obtained for
25 cycle pulses at an intensity of $2\times10^{13}\mbox{ }\mathrm{Wcm}^{-2}$,
giving, within an accuracy of 10\%, an asymmetry parameter $q=-1.84$ and line
width, $\Gamma=0.27$ eV.

The resonance parameters differ significantly from those obtained in previous
photoionization calculations \cite{taylor1975}. The present resonance width is
about a factor 4 larger than obtained previously, 0.068 eV. This is not too
surprising, since the present calculations are carried out only over a finite
laser pulse, which gives rise to a broadening of the resonance. The effects of
this broadening can be estimated by carrying out calculations for different
pulse lengths. Table \ref{tab:cons_int} indeed shows that the resonance width
decreases with increasing laser pulse length. In addition, the resonance may
also broaden due to the intensity of the laser pulse. At higher intensities
states can photoionize as well as autoionize. This leads to a reduction in the
lifetime of the state, and hence a broadening.  This effect is shown in Table
\ref{tab:cons_len}, where resonance parameters are compared as a function of
peak intensity.

Figure \ref{fig:res_prof} shows that at the fifth-harmonic stage, the
resonance has a strong effect on the harmonic yields.  The strong interference
between the continuum and the resonance leads to an asymmetric profile
with harmonic yields increased below and decreased above the resonance energy.
Both the continuum and the resonant pathway to HG must therefore be accounted
for. As these pathways involve the response of different electrons, it shows
that a multi-electron treatment of HG is necessary for Ar in
the 200-240nm wavelength range. 

Table \ref{tab:cons_int} shows that the asymmetry parameter is not affected
significantly by the pulse duration.  On the other hand, Table
\ref{tab:cons_len} shows that this parameter changes noticeably with the
intensity of the laser pulse. This knowledge is important for modelling the
harmonic response of a macroscopic system and the appearance of resonances over
a particular energy distribution.  However, it is difficult to analyze the
parameter changes in detail. At an intensity of $5\times10^{13}\mbox{
}\mathrm{Wcm}^{-2}$ the ponderomotive shift of the ground state means that at
the fifth-harmonic resonance, the third harmonic lies very close to the
$3s^23p^5$ threshold, and this may affect the profile. 

%%%%	Conclusions	%%%%
In conclusion we have applied TDRM theory to study the appearance of the
$3s3p^6n\ell$ autoionizing resonances in the fifth-harmonic for Ar. These
window resonances show significant asymmetry within the fifth-harmonic
spectrum, indicating interference between pathways involving the response of
different electrons.  This study provides a first demonstration of the new
capability within time-dependent R-matrix theory to obtain harmonic spectra in
multi-electron atoms with electron-electron correlation fully included. This
extension may enable further evaluation of whether and how evidence of
multi-electron dynamics can be found within harmonic spectra of atoms.

In the future, we hope to extend the approach to longer wavelengths and higher
intensities. This will, however, require an assessment of the different gauges
that can be used to obtain the harmonic spectrum, and extensive code
development. The TDRM approach has proven its merit for strong-field ioniztion
at wavelengths of 390nm at $4\times 10^{14}\mathrm{Wcm}^{-2}$ \cite{vdh07}. It will be of
interest to export the determination of harmonic yields to the newly developed
$R$-matrix with time-dependence codes \cite{RMT}, which may be better suited
for longer wavelengths.

ACB and SH acknowledge support from DEL NI under the programme for government.
HWH is supported by G/055816/1 from the ESPRC. 
MAL acknowledges funding from the Numerical Algorithms Group (NAG) Ltd.

\begin{center}
\begin{table}[h]
\caption{The asymmetry parameters, line widths and positions for the
resonance profiles due to $2\times10^{13}\mbox{ }\mathrm{Wcm}^{-2}$ laser
pulses of various lengths
accurate to within $\sim$10\% \label{tab:cons_int}}
\begin{tabular}{cccc}
& & &\\
\hline
\hline
Peak cycles & $q$ parameter & Line width (eV) & Position (eV)\\
\hline
10 & -1.83 & 0.50  & 26.99\\
15 & -1.82 & 0.40  & 27.01\\
20 & -1.83 & 0.30  & 27.00\\
25 & -1.84 & 0.27  & 27.01\\
\hline
\end{tabular}
\end{table}

\begin{table}[hb]
\caption{The asymmetry parameters, line widths and positions for the resonance
profiles due to 25 cycle laser pulses of various intensities, accurate to
within $\sim$10\% \label{tab:cons_len}}
\begin{tabular}{cccc}
& & & \\
\hline
\hline
Intensity  & $q$ parameter & Line width & Position \\
($10^{13}\mathrm{Wcm}^{-2}$) & & (eV) & (eV) \\
\hline
2.0  & -1.84 & 0.27 & 27.01 \\
3.0  & -1.79 & 0.30 & 27.08 \\
5.0  & -1.31 & 0.32 & 27.27 \\
\hline
\end{tabular}
\end{table}
\end{center}

\bibliography{mybib}

\end{document}